\documentclass[english,preprint,onecolumn,aps,pre,eqsecnum,longbibliography,showpacs,showkeys,a4paper]{revtex4-1}
\usepackage[latin9]{inputenc}
\usepackage{color}
\definecolor{note_fontcolor}{rgb}{0.80078125, 0.80078125, 0.80078125}
\usepackage{babel}
\usepackage{textcomp}
\usepackage{amsmath}
\usepackage{amssymb}
\usepackage{graphicx}
\usepackage{esint}
\usepackage[unicode=true,
 bookmarks=true,bookmarksnumbered=true,bookmarksopen=true,bookmarksopenlevel=1,
 breaklinks=true,pdfborder={0 0 0},backref=false,colorlinks=true]
 {hyperref}
\hypersetup{
 pdfauthor={AINS}}

\makeatletter

\newenvironment{lyxgreyedout}
  {\textcolor{note_fontcolor}\bgroup\ignorespaces}
  {\ignorespacesafterend\egroup}

\@ifundefined{textcolor}{}
{%
 \definecolor{BLACK}{gray}{0}
 \definecolor{WHITE}{gray}{1}
 \definecolor{RED}{rgb}{1,0,0}
 \definecolor{GREEN}{rgb}{0,1,0}
 \definecolor{BLUE}{rgb}{0,0,1}
 \definecolor{CYAN}{cmyk}{1,0,0,0}
 \definecolor{MAGENTA}{cmyk}{0,1,0,0}
 \definecolor{YELLOW}{cmyk}{0,0,1,0}
 }
\numberwithin{equation}{section}
\numberwithin{figure}{section}
\numberwithin{table}{section}

\usepackage{graphicx}

\DeclareGraphicsExtensions{.png,.pdf,.eps}
\graphicspath{{pictures/}}

\makeatother

\begin{document}

\title{The Quantum as an Emergent System}

\author{Gerhard \surname{Gr\"ossing}\textsuperscript{1}}

\email[E-mail: ]{ains@chello.at}

\homepage[Visit: ]{http://www.nonlinearstudies.at/}

\author{Siegfried \surname{Fussy}\textsuperscript{1}}

\email[E-mail: ]{ains@chello.at}

\homepage[Visit: ]{http://www.nonlinearstudies.at/}

\author{Johannes \surname{Mesa Pascasio}\textsuperscript{1,2}}

\email[E-mail: ]{ains@chello.at}

\homepage[Visit: ]{http://www.nonlinearstudies.at/}

\author{Herbert \surname{Schwabl}\textsuperscript{1}}

\email[E-mail: ]{ains@chello.at}

\homepage[Visit: ]{http://www.nonlinearstudies.at/}

\affiliation{\textsuperscript{1}Austrian Institute for Nonlinear Studies, Akademiehof\\
 Friedrichstr.~10, 1010 Vienna, Austria}

\affiliation{\textsuperscript{2}Institute for Atomic and Subatomic Physics, Vienna
University of Technology\\
Operng.~9, 1040 Vienna, Austria\vspace*{2cm}
}

\date{\today}
\begin{abstract}
Double slit interference is explained with the aid of what we call
``21\textsuperscript{st}century classical physics''. We model a
particle as an oscillator (``bouncer'') in a thermal context, which
is given by some assumed ``zero-point'' field of the vacuum. In
this way, the quantum is understood as an \emph{emergent system},
i.e., a steady-state system maintained by a constant throughput of
(vacuum) energy. To account for the particle's thermal environment,
we introduce a ``path excitation field'', which derives from the
thermodynamics of the zero-point vacuum and which represents all possible
paths a particle can take via thermal path fluctuations. The intensity
distribution on a screen behind a double slit is calculated, as well
as the corresponding trajectories and the probability density current.
Further, particular features of the relative phase are shown to be
responsible for nonlocal effects not only in ordinary quantum theory,
but also in our classical approach.%
\begin{lyxgreyedout}
\global\long\def\VEC#1{\mathbf{#1}}

\global\long\def\d{\,\mathrm{d}}

\global\long\def\e{{\rm e}}

\global\long\def\meant#1{\left<#1\right>}

\global\long\def\meanx#1{\overline{#1}}

\global\long\def\mpbracket{\ensuremath{\genfrac{}{}{0pt}{1}{-}{\scriptstyle (\kern-1pt +\kern-1pt )}}}

\global\long\def\pmbracket{\ensuremath{\genfrac{}{}{0pt}{1}{+}{\scriptstyle (\kern-1pt -\kern-1pt )}}}

\global\long\def\p{\partial}
\end{lyxgreyedout}

\end{abstract}


\maketitle

\section{Introduction: ``Natural explanations'' and emergence\label{sec:intro}}

In his book review of Steve Adler's \emph{Quantum Theory as an Emergent
Phenomenon} \cite{Adler.2004quantum}, Philip Pearle illustrated the
concept of an ``enlightenment task'' by ``...trying to explain
the unnatural by the natural \textminus{} in this case, the \textquoteleft{}unnatural\textquoteright{}
being quantum physics and the \textquoteleft{}natural\textquoteright{}
being classical physics\ldots{}\textquotedblright{} \cite{Pearle.2005stephen}.
Such tasks were already undertaken in the formative years of quantum
theory, and Pearle characterized them very aptly in the following
way:\smallskip{}

\begin{minipage}[t]{0.95\columnwidth}%
``Classical particles and their dynamics are re-introduced, but a
strong element of the unnatural remains. In the deBroglie-Bohm and
Madelung models, it is the mysterious quantum force. In the Nelson
model, it is the mysterious backward diffusion process (which, together
with the usual classical forward diffusion process, forces a particle\textquoteright{}s
drift \textendash{} its mean position \textendash{} to be a dynamically
determined quantity instead of, as classically, an independent variable
set by external influences).''%
\end{minipage}\medskip{}

Now, the ``mysterious quantum force'' ultimately derives from the
``mysterious'' wave function $\Psi$, a solution of the Schr\"odinger
equation, which is an initial ingredient for any Bohmian-type model
\cite{Bohm.1952interpr1,Bohm.1993undivided,Holland.1993}. Whereas
in the latter the existence of $\Psi$ is usually not questioned and
thus remains ``unnatural'' in Pearle's terms, the Nelsonian model
actually is an attempt to rely on Newtonian physics only \cite{Nelson.1966derivation,Nelson.1985quantum,Fritsche.2003new},
or on what one may thus also call ``natural explanations''. Pearle's
comment apparently leaves two possibilities to overcome the ``mysterious''
element of the Nelson model, i.e., of backward diffusion. One would
be to substitute the combination of forward and backward diffusion
processes by some other process where a particle's drift would turn
out as a simple classical variable determined by external influences.
As this is not viable since, upon the introduction of any usual forward
diffusion, one does need a balancing process in opposition to it,
the second possibility remains that a particle's drift actually \emph{is}
a dynamically determined quantity, albeit in a new framework. This
framework, we propose, would have to be some kind of steady-state
maintained by a throughput of energy from the (``contextual'') environment.

In our ``sub-quantum thermodynamics'' approach to model quantum
systems \cite{Groessing.2008vacuum,Groessing.2009origin,Groessing.2010emergence,Groessing.2010entropy,Groessing.2011dice,Groessing.2011explan,Groessing.2012doubleslit},
we pursue the last option as a concrete possibility: not only do we
consider \emph{quantum theory} as emergent, but also, more specifically,
the \emph{quantum systems} themselves. This means that we refrain
from any attempt to model the quantum on some singular, ``basic''
level only, but rather consider it as a ``self-organizing system''
whose description requires a more encompassing combination of levels.
One of the first definitions in science of such a system was given
by Heinz von Foerster, i.e., as ``a system which maps order of its
environment onto its own organization.''\cite{von_Foerster.1960self-organizing}
From the 1960ies onward, an ever increasing number of studies has
been published, in a variety of different disciplines, on self-organization,
and later on emergence, respectively. This tendency finally also entered
the discussions on the foundations of physics, notably in the works
by Ilya Prigogine \cite{Prigogine.1962non-equilibrium}, or in famous
essays by P. W. Anderson (``More is Different'' \cite{Anderson.1972more}),
and later by Sam Schweber \cite{Schweber.1993physics}, as well as
recently in popularized form by Robert B. Laughlin \cite{Laughlin.2006different}.
In a slight shift of emphasis from self-organization to emergence,
Sam Schweber characterized the latter in the following way:\smallskip{}

\begin{minipage}[t]{0.95\columnwidth}%
\textquotedblleft{}It is not enough to know the \textquoteleft{}fundamental\textquoteright{}
laws at a given level. It is the solutions to equations, not the equations
themselves, that provide a mathematical description of the physical
phenomena. \textquoteleft{}Emergence\textquoteright{} refers to properties
of the solutions \textendash{} in particular, the properties that
are not readily apparent from the equations.\textquotedblright{} \cite{Schweber.1993physics}%
\end{minipage}\medskip{}

Schweber thus refers to the possibility that the ``fundamental laws''
on some ``basic level'' may by themselves not always be sufficient
to grasp the essence of some physical phenomenon. This is reflected
also in our group's approach where the quantum is modeled as a self-organizing,
\emph{dynamical} entity whose complete description needs more than
just one (presumably ``basic'') level. As mentioned, we thus propose\emph{
that a quantum system be considered as a well-coordinated emergent
system}. In doing so, we consider particle-like and wave-like phenomena
as the result of both stochastic and regular dynamical processes.
A prototype of such a system is well known from classical physics,
viz., the ``bouncing droplets'' of Couder's group \cite{Couder.2005,Couder.2006single-particle,Protiere.2006,Eddi.2009,Fort.2010path-memory},
which in fact exhibit a whole series of phenomena reminiscent of quantum
ones. Analogously, our group has in recent years attempted to model
a quantum as a non-equilibrium steady-state maintained by a permanent
throughput of energy. Specifically, we consider a ``particle'' as
a ``bouncer'' whose oscillations are phase-locked with those of
the energy-momentum reservoir of the surrounding ``vacuum'', i.e.,
the zero-point field. (Note that the possible existence of a corresponding,
underlying ``medium'' is \textit{a priori} independent of quantum
theory. For a similar view, compare, for example, the approach of
``stochastic electrodynamics'' by Cetto and de~la~Pe\~na~\cite{Pena.1996,Cetto.2012quantization}.
For similar approaches in terms of assuming some sub-quantum, thermodynamic
or hybrid-type variant of emergent quantum physics, see \cite{Adler.2004quantum,Acosta.2011entropic,Elze.2011general,Garbaczewski.1992derivation,Khrennikov.2011prequantum,t_Hooft.2007emergent,Wetterich.2009zwitters:,Adler.2012quantum,Carroll.2012remarks,Elze.2012four,Faber.2012particles,Garbaczewski.2012probabilistic,Hofer.2012extended,t_Hooft.2012quantum,Acosta.2012holographic,Khrennikov.2012classical,Nelson.2012review,Ord.2012quantum,Jizba.2012quantum,Schuch.2012nonlinear,Wetterich.2012quantum})

In other words, we attempt to model the quantum in a ``classical''
framework. Note, however, that we consider this framework more specifically
to comprise what one may call ``21\textsuperscript{st} century classical
physics'', i.e., including all the recent developments in the fields
of nonequilibrium thermodynamics, ballistic diffusion, diffusion wave
fields, and the like. More concretely, in assuming that (part of)
the ``classical'' zero-point fluctuations undergo regular oscillatory
motion, where the latter is partly caused by and dynamically coupled
to the oscillator's frequency $\omega$, we are able to derive fundamental
elements of quantum theory from a purely classical approach. In Ref.~\cite{Groessing.2011explan},
as well as in Schwabl~\cite{Schwabl.2012quantum}, we have shown
how Planck's relation between the energy $E$ and the frequency $\omega$,
$E=\hbar\omega$, can be derived from a sub-quantum physics, with
Planck's (reduced) constant $\hbar$ indicating a universal angular
momentum, and we have also shown that with this relation alone one
can derive the exact Schr\"odinger equation from (modern) classical
physics \cite{Groessing.2008vacuum,Groessing.2009origin}. Moreover,
also the stochastic element of the zero-point fluctuations enters
decisively into our model, such that in effect we obtain an exact
description of free quantum motion via a combination of the propagation
of classical Huygens-type waves with diffusion due to stochastic sub-quantum
mechanics \cite{Groessing.2010emergence}. We particularly stress
that the ``particle'' is considered as an off-equilibrium steady
state oscillator maintained by a constant throughput of energy provided
by the zero-point field. Thus, it is exactly this \emph{energy throughput}
which is responsible for a particle's \emph{natural drift} in Pearle's
sense: it is the permanent absorption and re-emission of kinetic energy
which will be explicated below to provide a ``natural'' diffusion
model, albeit one of a specific kind, which is called ``ballistic
diffusion''. So, a quantum in our model emerges from the synchronized
dynamical coupling between an oscillator (``bouncer'') and its wave-like
environment. In sum, with this ansatz, we have been able to derive
from ``21\textsuperscript{st} century classical physics'' the following
quantum mechanical features:
\begin{itemize}
\item Planck\textquoteright{}s relation for the energy of a particle, 
\item the exact (one- and n-particle) Schr\"odinger equation for conservative
and non-conservative systems, 
\item the Heisenberg uncertainty relations, 
\item the quantum mechanical superposition principle and Born\textquoteright{}s
rule, 
\item the quantum mechanical decay of a Gaussian wave packet, 
\item quantum mechanical interference at the double slit.
\end{itemize}
Although in our model Huygens' principle applies, it does so only
for an idealized combination of a ``walking'' motion of some velocity
$\VEC v$ with the centrally-symmetric diffusion waves' motions of
velocity $\VEC u$ orthogonal to it. Effectively, it only holds approximately,
disturbed by that part of the accompanying diffusive process which
is to be described by the corresponding velocity fluctuations $\delta\VEC u$.
In fact, as we have shown for a Gaussian slit, the exact quantum mechanical
result can be described as a combination of classical wave mechanics
with the addition of a corresponding stochastic diffusion process
\cite{Groessing.2010emergence}. We have also shown \cite{Groessing.2012doubleslit}
that the same modelling procedure also perfectly applies to a double
slit system. In particular, to make our point as clear as possible,
we provide here a more extensive discussion of a simple calculational
tool related to what we call the ``path excitation field''. With
it, one can easily derive results for quantum mechanics without ever
using complex-valued functions such as wave functions, for example.

\section{The path excitation field: A classical explanatory framework for
Gaussian dispersion and double-slit interference\label{sec:path}}

To begin with, we recall some of the basic results of our earlier
work, including that on diffraction at a single Gaussian slit \cite{Groessing.2010emergence}.
We claim for a particle of frequency $\omega$ embedded in a stochastic
(``zero-point'') environment that its \textit{average total energy}
is given by the average ``total'' energy $\meanx{\hbar\omega}$
of the particle itself plus a kinetic energy term due to momentum
changes $\VEC p_{u}=:m\VEC u$ which it receives from or gives off
to the environment: 
\begin{equation}
\meanx{E_{{\rm tot}}}=\meanx{\hbar\omega}+\frac{\meanx{p_{u}^{2}}}{2m}=\text{const},\label{eq:2.1}
\end{equation}
 where the averaging (as denoted by the bars) is defined in $n$-dimensional
configuration space as 
\begin{equation}
\meanx{p_{u}^{2}}:=\int Pp_{u}^{2}\d^{n}x.\label{eq:2.3}
\end{equation}
 $P=P(\VEC x,t)$ refers to the probability density of some relevant
distribution. For our model system, the latter is given as a solution
of a generalized (``anomalous'') diffusion equation, i.e., with
a time-dependent diffusion coefficient. As can be seen from \eqref{eq:2.1},
the momentum changes can be either positive or negative, and actually
will \textit{on the average} balance each other, since they are \textit{a
priori} unbiased. The deeper reason for this balancing in our model
is due to the fact that we consider the quantum to be a steady-state
system in the sense that its ``total'' energy $\hbar\omega$ is
maintained over times $t\gg1/\omega$ by the permanent throughput
of kinetic energy $p_{u}^{2}/2m$. In other words, to maintain the
steady-state, during the intervals of the average order of $t\simeq1/\omega$,
there will both be an absorption of a momentum $p_{u}=mu$ and a release
of the same amount, $p_{u}=-mu$, thus providing a ``natural'' explanation
of the involved diffusion processes as envisaged in the introduction.

In our earlier papers \cite{Groessing.2008vacuum,Groessing.2009origin,Groessing.2010emergence},
we have shown that, apart from the ordinary particle current $\VEC J(\VEC x,t)=P(\VEC x,t)\VEC v$,
we are thus dealing with two additional, yet opposing, currents $\VEC J_{u}=P(\VEC x,t)\VEC u$,
which are on average orthogonal to $\VEC J$ \cite{Groessing.2008vacuum,Groessing.2009origin,Groessing.2010emergence,Groessing.2011explan},
and which are the emergent outcome from the presence of numerous corresponding
velocities 
\begin{equation}
\VEC u_{\pm}=\mp\frac{\hbar}{2m}\frac{\nabla P}{P}.\label{eq:2.7}
\end{equation}

We denote with $\VEC u_{+}$ and $\VEC u_{-}$, respectively, the
two opposing tendencies of the diffusion process. In the reference
frame of a single free particle, and starting at $t=0$ at the center
of the distribution $P$, the averages obey 
\begin{equation}
\meanx{\VEC u}_{-}(\VEC x,t)=-\meanx{\VEC u}_{+}(-\VEC x,t).\label{eq:2.7a}
\end{equation}

Now let us consider an experimental setup with a particle source.
To describe the velocity distribution, we introduce a velocity field
with average velocity $\meanx{\VEC v}$, and amplitudes $R(\VEC x,t)$.
As mentioned, we refer to their intensities $P=R^{2}$ as the solutions
of a diffusion equation. These typically appear in the form of a Gaussian
distribution $P(\VEC x,t)$ of possible particle locations $\VEC x$,
even if there is only one particle at a time emerging from the corresponding
``Gaussian slit'', in one dimension for simplicity, 
\begin{equation}
P(x,t)=\frac{1}{\sqrt{2\pi}\,\sigma}\e^{-\frac{(x-x_{0})^{2}}{2\sigma^{2}}},\label{eq:2.8}
\end{equation}
 with the usual variance $\sigma^{2}=\meanx{\left(\Delta x\right)^{2}}=\meanx{\left(x-x_{0}\right)^{2}}$,
and where we choose $x_{0}=0$ furtheron. Regarding $\VEC u$, even
in this scenario of one-particle-at-a-time, we deal with an \textit{ensemble}
of velocity vectors $\VEC u_{\alpha}(t)$ representing hypothetical
motions on the sub-quantum level in a small volume around $\VEC x$,
whose mean value will be given by 
\begin{equation}
\VEC u(\VEC x,t)=\frac{1}{N(\VEC x,t)}\sum_{\alpha=1}^{N(\VEC x,t)}\VEC u_{\alpha}(t).\label{eq:2.9}
\end{equation}
 Here, the (typically very large) number $N$ refers to the number
of possible path directions of the bouncer due to the existence of
the wave-like excitations of the zero-point field. This may be reminiscent
of Feynman's picture of photons \textit{virtually} probing every possible
path in an experimental setup, but in our case it is a configuration
of \textit{real} wave-like excitations, which in a resulting Brownian-type
motion guide the bouncer along its path of average velocity $\VEC u$.
Again, we note that here we discuss not only a passive guidance of
the ``particle'' by the surrounding wave configurations, but point
out also the very active role of the ``particle'' in (partly) \textit{creating}
said wave configurations due to the effects of its bouncing. To account
for \eqref{eq:2.7}, we split up $\VEC u(\VEC x,t)$ according to
\begin{equation}
\VEC u(\VEC x,t)=\frac{1}{2N}\left[\sum_{\alpha=1}^{N}\VEC u_{\alpha,+}+\sum_{\alpha=1}^{N}\VEC u_{\alpha,-}\right]=\frac{1}{2}\left[\VEC u_{+}+\VEC u_{-}\right],\label{eq:3.6}
\end{equation}
 thus reflecting the isotropy of the diffusion process. Still, the
uncontrollable and possibly unknowable velocity field $\VEC u$ representing
the Brownian motion of the bouncer may not be operational, but when
we take the average according to the rule \eqref{eq:2.3}, we obtain
a ``smoothed-out'' \textit{average velocity field} 
\begin{equation}
\meanx{\VEC u(\VEC x,t)}=\int P\VEC u(\VEC x,t)\d^{n}x,\label{eq:2.10}
\end{equation}
 which is all that we need for our further considerations. Similarly,
based on the fact that we have an initial Gaussian distribution of
velocity vectors $\VEC v(\VEC x,t)$, we define an average velocity
field $\meanx{\VEC v}$ of the wave propagation as 
\begin{equation}
\meanx{\VEC v(\VEC x,t)}=\int P\VEC v(\VEC x,t)\d^{n}x,\label{eq:2.11}
\end{equation}
and make use of an average orthogonality between the two velocity
fields, $\VEC u$ and $\VEC v$, \cite{Groessing.2008vacuum,Groessing.2009origin,Groessing.2010emergence,Groessing.2011dice},
\begin{equation}
\meanx{\VEC v\cdot\VEC u}=\int P\VEC v\cdot\VEC u\d^{n}x=0.\label{eq:2.12}
\end{equation}

In effect, then, the combined presence of both velocity fields $\VEC u$
and $\VEC v$ can be denoted as a \textit{path excitation field}:
via diffusion, the bouncer in its interaction with already existing
wave-like excitations of the environment creates an ``agitated'',
or ``heated-up'', thermal ``landscape'', which can also be pictured
by interacting wave configurations all along between source and detector
of an experimental setup. Recall that our prototype of a ``walking
bouncer'', i.e., from the experiments of Couder's group, is always
driven by its interactions with a superposition of waves emitted at
the points it visited in the past. Couder et~al. denote this superposition
of in-phase waves the ``path memory'' of the bouncer \cite{Fort.2010path-memory}.
This implies, however, that the bouncers at the points visited in
``the present'' necessarily create new wave configurations which
will form the basis of a ``path memory'' in the future. In other
words, the wave configurations of the past determine the bouncer's
path in the present, whereas its bounces in the present co-determine
the wave configurations at any of the possible locations it will visit
in the future. Therefore, we call the latter configurations the \textit{path
excitation field}, which may also be described as ``heated-up''
thermal field. As in the coupling of an oscillator with classical
diffusion, diffusion wave fields arise with instantaneous field propagation
\cite{Groessing.2009origin,Mandelis.2001structure}, one has elements
of the whole setup which may be nonlocally oscillating (``breathing'')
in phase. This means that the Gaussian of \eqref{eq:2.8} does represent
a nonlocal path excitation field in that it is a physically existing
and effective entity responsible for where the bouncing ``particle''
can possibly go. As we have shown \cite{Groessing.2010emergence},
one can in this classical framework, along with the time-dependence
of the diffusivity, effectively and easily describe the (sub-)quantum
physics of diffraction at a Gaussian slit, which we now briefly recapitulate.

At first we note that Eq.~\eqref{eq:2.1} is an \textit{average}
energy conservation law only. This means that apart from the momentum
changes $\VEC p_{u}=\pm m\VEC u$ discussed so far, also variations
in $\VEC p_{u}$ will have to be taken into account, and thus also
variations in the ``particle energy'' $\hbar\omega$. If for the
latter one just considers its kinetic energy term, $mv^{2}/2$, then
said variations will lead to exchanges of velocity/momentum terms
providing the net balance 
\begin{equation}
m\delta\VEC v=m\delta\VEC u.\label{eq:2.13}
\end{equation}
Using the expression \eqref{eq:2.7} for $\VEC u$, one obtains with
the Gaussian~\eqref{eq:2.8}, $\overline{\VEC p_{u}^{2}}\mid_{t=0}=:m^{2}u_{0}^{2}$,
and with $\meanx{(\nabla\ln P)^{2}}=-\meanx{\nabla^{2}\ln P}$, that
\begin{equation}
u_{0}^{2}=\frac{D^{2}}{\sigma_{0}^{2}}=\meanx{u^{2}}+\meanx{(\delta u)^{2}}=\frac{D^{2}}{\sigma^{2}}+\meanx{(\delta u)^{2}},\label{eq:2.17}
\end{equation}
 where as usual $\sigma=\sigma(t)=\sqrt{\,\meanx{x^{2}}}$ for $x_{0}(t=0)=0$,
and $\sigma_{0}=\sigma(t=0)$. One can view the Gaussian distribution
$P$ of kinetic energy also as a sort of ``heat accumulation'',
which has its maximum at the center. Considering now the application
of momentum fluctuations (up to second order) to a particle with initial
($t=0$) distance $x(0)$ from said center, with the fluctuation term
for $t>0$ defined as $\VEC p_{u}\pm\delta\VEC p_{u}=\pm m(\VEC u\pm\delta\VEC u)$,
one obtains at time $t$ the envisaged ``natural'' drift as 
\begin{equation}
\VEC x(t)=\VEC x(0)\pm\left(\VEC u\pm\delta\VEC u\right)t.\label{eq:2.18}
\end{equation}
 Squaring Eq.~\eqref{eq:2.18} and forming the r.m.s.\ can then
easily be shown to provide \cite{Groessing.2010emergence}, 
\begin{equation}
\meanx{x^{2}}(t)=\meanx{x^{2}}(0)+\left[\meanx{u^{2}}+\meanx{(\delta u)^{2}}\right]t^{2}=\meanx{x^{2}}(0)+u_{0}^{2}t^{2}.\label{eq:2.19}
\end{equation}
Comparing with Eq.~\eqref{eq:2.17} also provides the time evolution
of the wave packet's variance as 
\begin{equation}
\sigma^{2}=\sigma_{0}^{2}\left(1+\frac{D^{2}t^{2}}{\sigma_{0}^{4}}\right),\label{eq:2.20}
\end{equation}
and finally the average velocity field of a Gaussian wave packet as
\begin{equation}
v_{{\rm tot}}(t)=v(t)+\left[x(t)\right]\frac{u_{0}^{2}t}{\sigma^{2}}.\label{eq:2.23}
\end{equation}

Note that Eqs.~\eqref{eq:2.20} and \eqref{eq:2.23} are derived
solely from classical physics. Still, they are in full accordance
with quantum theory, and in particular with Bohmian trajectories \cite{Holland.1993}.
Note also that one can rewrite Eq.~\eqref{eq:2.19} such that it
appears like a linear-in-time formula for Brownian motion, 
\begin{equation}
\meanx{x^{2}}=\meanx{x^{2}(0)}+D(t)\, t,\label{eq:2.24}
\end{equation}
 where a time dependent diffusivity 
\begin{equation}
D(t)=u_{0}^{2}\, t=\frac{\hbar^{2}}{4m^{2}\sigma_{0}^{2}}\, t\label{eq:2.25}
\end{equation}
 characterizes Eq.~\eqref{eq:2.24} as \textit{ballistic diffusion}.
The appearance of a time-dependent $D(t)$ is essential, but also
straightforward. The diffusivity is changed over time, because the
``particle's'' thermal environment changes: With the ``heat''
initially concentrated within the narrow spatial constraints determined
by $\sigma_{0}$ of the source (``Gaussian slit''), $D(t)$ must
become larger with time because of the gradually lower heat concentration
due to dissipation into the unconstrained environment. (Note that
a similar scenario was suggested by Garbaczewski~\cite{Garbaczewski.1992derivation};
others are presently intensively discussed in the context of the so-called
``superstatistics'' \cite{Beck.2008}.) This makes it possible to
simulate the dispersion of a Gaussian wave packet on a computer by
simply employing coupled map lattices for classical diffusion, with
the diffusivity given by Eq.~\eqref{eq:2.25}. (For detailed discussions,
see refs.~\cite{Groessing.2010emergence} and \cite{Groessing.2011dice}.)

With the essentials of Gaussian dispersion at our disposal, it is
very simple to now also describe and explain quantum interference
with our approach. \cite{Groessing.2012doubleslit} We have chosen
a textbook scenario in the form of the calculation of the intensity
distribution and the particle trajectories in an electron interferometer.
As we are also interested in the trajectories, we refer to, and compare
our results with, the well-known work by Phillipidis et~al. \cite{Philippidis.1979quantum},
albeit in the form as presented by Holland \cite{Holland.1993}.

We choose similar initial situations as in \cite{Holland.1993}, i.e.,
electrons (represented by plane waves in the forward $y$-direction)
from a source passing through ``soft-edged'' slits $1$ and $2$
in a barrier (located along the $x$-axis) and recorded at a screen.
In our model, we therefore note two Gaussians representing the totality
of the effectively ``heated-up'' path excitation field, one for
slit $1$ and one for slit $2$, whose centers have the distances
$+X$ and $-X$ from the plane spanned by the source and the center
of the barrier along the $y$-axis, respectively. 

As it is well known from classical wave mechanics, the total amplitude
$R$ of two coherent waves with (suitably normalized) amplitudes $R_{i}=\sqrt{P_{i}}$,
$i=1$ or $2$, is given by
\begin{align}
R\left(\VEC{\VEC k_{1},\VEC k_{2},r}\right) & =R_{1}\cos\left(\omega t-\VEC k_{1}\cdot\VEC r+\varphi_{0}\right)+R_{2}\cos\left(\omega t-\VEC k_{2}\cdot\VEC r+\varphi_{0}\right)\nonumber \\
 & =R_{1}\cos\left(\omega t-\varphi_{1}\right)+R_{2}\cos\left(\omega t-\varphi_{2}\right),
\end{align}
where $\varphi_{0}$ is some initial phase at $t=0$, and $\varphi_{i}=\VEC k_{i}\cdot\VEC r-\varphi_{0}$
. Considering the average of this expression over a multitude of similar
wave superpositions, one obtains as usual the \emph{averaged total
intensity}
\begin{equation}
P_{{\rm tot}}:=R^{2}=R_{1}^{2}+R_{2}^{2}+2R_{1}R_{2}\cos\varphi=P_{1}+P_{2}+2\sqrt{P_{1}P_{2}}\cos\varphi,\label{eq:3.0a}
\end{equation}
where $\varphi=\varphi_{1}-\varphi_{2}=\left(\VEC k_{1}-\VEC k_{2}\right)\cdot\VEC r$.
Note that the relative phase difference $\varphi$ enters Eq.~\eqref{eq:3.0a}
only via the cosine function, such that, e.g., even if the total wave
numbers (and thus also the total momenta) $\VEC k_{i}$ were of vastly
different size, the cosine effectively makes Eq.~\eqref{eq:3.0a}
independent of said sizes, but dependent only on an angle modulo $2\pi$.
This will turn out as essential for our discussion further below.

Now, the $x$-components of the centroids' motions from the two alternative
slits $1$ and $2$, respectively, are given by the ``particle''
velocity components 
\begin{equation}
v_{x}=\pm\frac{\hbar}{m}\, k_{x},\label{eq:3.1}
\end{equation}
 respectively, such that the relative group velocity of the Gaussians
spreading into each other is given by $\Delta v_{x}=2v_{x}$. However,
in order to calculate the phase difference $\varphi$ descriptive
of the interference term of the intensity distribution \eqref{eq:3.0a},
one must take into account the total momenta involved, i.e, one must
also include the wave packet dispersion as described in the previous
Chapter. Thus, one obtains with the displacement $\pm x\left(t\right)=\mp\left(X+v_{x}t\right)$
in Eq.~\eqref{eq:2.23} the total relative velocity of the two Gaussians
as 
\begin{equation}
\Delta v_{{\rm tot},x}=2\left[v_{x}-(X+v_{x}t)\frac{u_{0}^{2}t}{\sigma^{2}}\right].\label{eq:3.1a}
\end{equation}
Therefore, the total phase difference between the two possible paths
(i.e., through either slit) becomes 
\begin{equation}
\varphi=\frac{1}{\hbar}(m\Delta v_{{\rm tot},x}\, x)=2mv_{x}\frac{x}{\hbar}-(X+v_{x}t)x\frac{1}{D}\frac{u_{0}^{2}t}{\sigma^{2}}.\label{eq:3.2}
\end{equation}
The Gaussians $P_{1}$ and $P_{2}$ for the corresponding slits are
given as 
\begin{equation}
P_{1}(x,t)=\frac{1}{\sqrt{2\pi\sigma^{2}}}\e^{-[x-(X+v_{x}t)]^{2}/2\sigma^{2}},\label{eq:3.3b}
\end{equation}
 and 
\begin{equation}
P_{2}(x,t)=\frac{1}{\sqrt{2\pi\sigma^{2}}}\e^{-[x+(X+v_{x}t)]^{2}/2\sigma^{2}}.\label{eq:3.3c}
\end{equation}
With equal amplitudes $R=R_{i}=\sqrt{P_{i}}$, for $i=1,2$, of the
Gaussians, and with normalization constant $N$, we thus obtain the
usual interference pattern in the form of the intensity distribution:
\begin{equation}
P_{{\rm tot}}(x,t)=R^{2}N^{2}\frac{1}{\sqrt{2\pi\sigma^{2}}}\e^{-[x^{2}+(X+v_{x}t)^{2}]/2\sigma^{2}}\left\{ e^{x(X+v_{x}t)/\sigma^{2}}+e^{-x(X+v_{x}t)/\sigma^{2}}+2\cos\varphi\right\} ,\label{eq:3.4}
\end{equation}
with the relative phase $\varphi$ given by Eq.~\eqref{eq:3.2}.
Whereas the first exponential describes the enveloping Gaussian, the
term inside the curly brackets of Eq.~\eqref{eq:3.4} describes the
interference fringes whose ``dark'' nodes are at the locations well-known
from textbooks, 
\begin{equation}
x=(n+\frac{1}{2})\pi/k_{x},\label{eq:3.5}
\end{equation}
if $X=-v_{x}t$, as the wave packets must approach each other ($v_{x}<0$)
for $t>0$. 

\begin{figure}
\centering{}\includegraphics[angle=90,width=0.85\columnwidth]{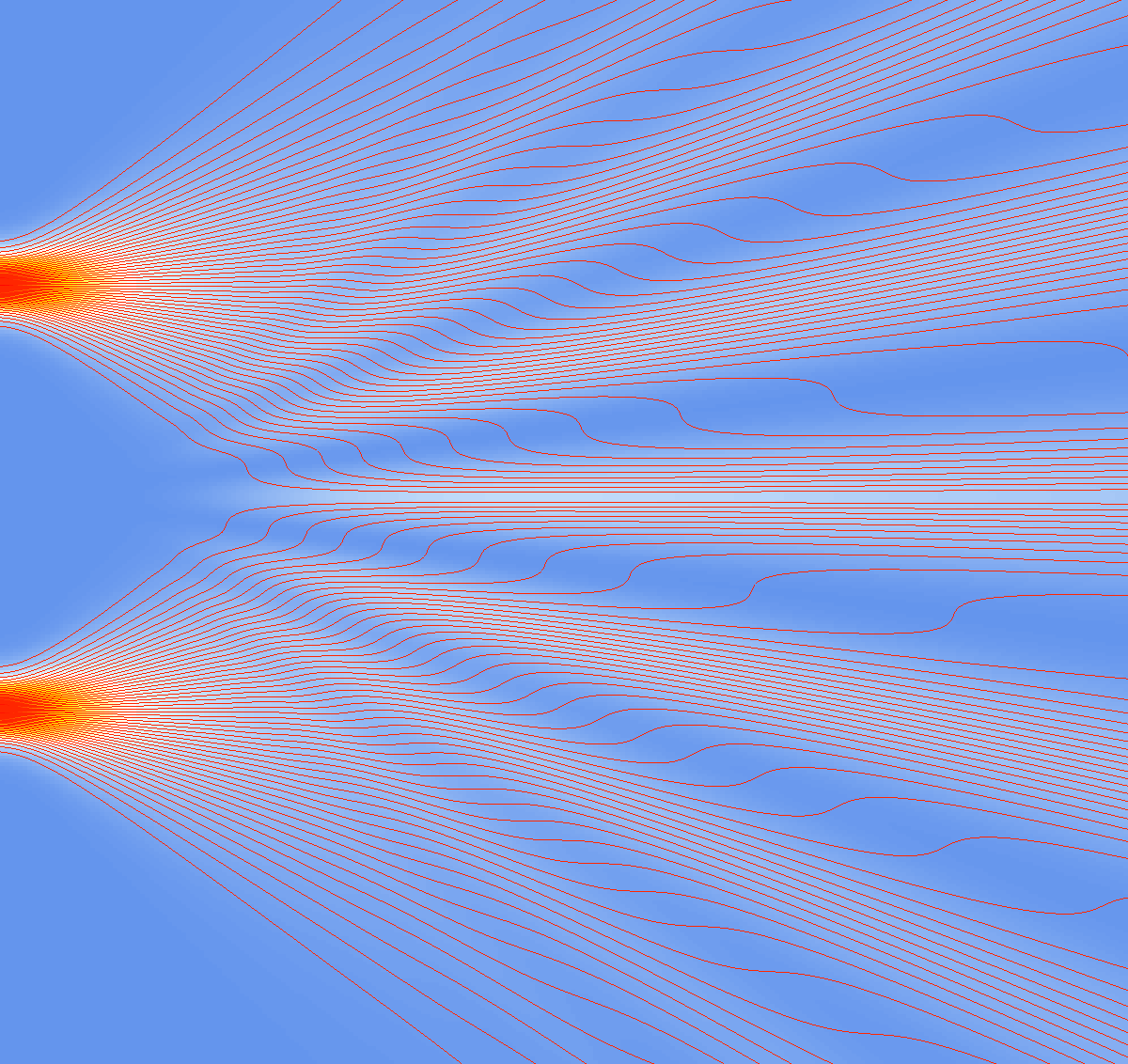}
\caption{Classical computer simulation of the interference pattern: intensity
distribution with increasing intensity from white through yellow and
orange, with averaged trajectories (red) for two Gaussian slits, and
with large dispersion (evolution from bottom to top; $\overline{v_{x,1}}=\overline{v_{x,2}}=0$).
The interference hyperbolas for the maxima characterize the regions
where the phase difference $\varphi=2n\pi$, and those with the minima
lie at $\varphi=(2n+1)\pi$, $n=0,1,2,\ldots$ Note in particular
the ``kinks'' of trajectories moving from the center-oriented side
of one relative maximum to cross over to join more central (relative)
maxima. In our classical explanation of interference, a detailed micro-causal
account of the corresponding kinematics can be given. The averaged
trajectories follow a Bohm-type ``no crossing'' rule: particles
from the left slit stay on the left side and \textit{vice versa} for
the right slit. This feature is explained here by a sub-quantum build-up
of kinetic (heat) energy acting as an emergent repellor along the
symmetry line.\label{fig:2}}
\end{figure}

Fig.~\ref{fig:2} depicts the interference of two beams emerging
from Gaussian slits established with the aid of a purely classical
simulation. As in the case of a Gaussian slit \cite{Groessing.2010emergence},
we again simulate diffusion with a time-dependent diffusivity $D(t)$.
To account for interference, we simply follow the classical rule for
the intensities \eqref{eq:3.0a}, with $\varphi$ from Eq.~\eqref{eq:3.2}.
The averaged trajectories are the flux lines obtained by choosing
a set of equidistant initial points at $y=0$. Two adjacent flux lines
thereby define regions of constant flux, i.e., $\int_{A}P\d A=\mathrm{const.}$,
with $A$ being the cross section of a flux tube.

Just as with the dispersion of a single Gaussian, we want to stress
that also the quantum interference pattern \eqref{eq:3.4} has been
derived here without the use of quantum mechanics, but solely on the
basis of classical physics. Exploiting our concept of the ``path
excitation field'', we have for this derivation implicitly used the
assumption of velocity fields $\VEC u(\VEC x,t)$ and $\VEC v(\VEC x,t)$,
respectively, which have entered the expression of a Gaussian's average
velocity field, Eq.~\eqref{eq:2.23}, and which in turn were shown
to essentially contribute to the interference pattern \eqref{eq:3.4}.
In what follows, we now want to make the use of these velocity fields
more explicit, i.e., we shall now concentrate on understanding the
emerging particle trajectories during quantum interference on the
basis of our classical velocity fields. As it turns out, a thorough
consideration of the nature of the relative phase $\varphi$ will
make it possible to obtain a deeper understanding of quantum interference
in general.

\section{Geometric meaning of the path excitation field\label{sec:geometric}}

Let us consider a single, classical ``particle'' (``bouncer'')
following the propagation of a set of waves of equal amplitude $R_{i}$,
each representing one of $i$ possible alternatives according to our
principle of path excitation. We first note that for the superposition
of two weighted wave vectors, with resultant vector $\VEC k$ and
total amplitude $R_{\mathrm{tot}}$, 
\begin{equation}
R_{\mathrm{tot}}\VEC k=R_{1}\VEC k_{1}+R_{2}\VEC k_{2},\label{eq:3.0-1}
\end{equation}
where the \emph{averaged scalar product} of the two associated unit
vectors is given by $\VEC{\hat{k}}_{1}\cdot\VEC{\hat{k}}_{2}$= cos$\varphi$
in accordance with \eqref{eq:3.0a}. 

\begin{figure}
\centering{}
\includegraphics[width=117mm,height=105mm]{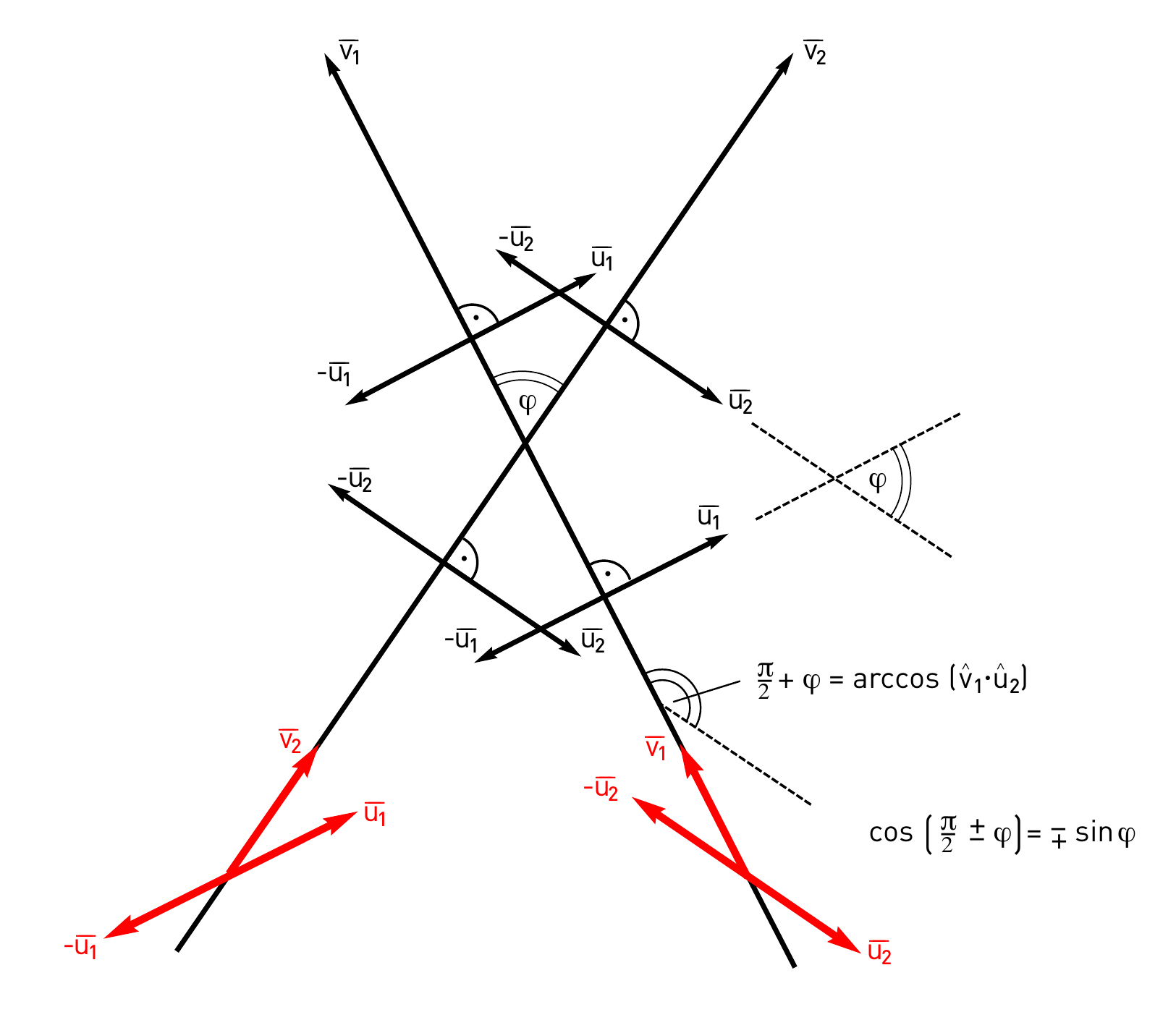}
\caption{Schematic of the phase angles between various components of the path
excitation field behind a double slit. $\overline{u_{i}}$ and $\overline{v_{i}}$,
$i=1$ or $2$, denote the average velocity fields involved, with
$\overline{v_{i}}$ referring to single free ``particle'' velocities
(i.e., either through slit 1 or slit 2), and the $\overline{u_{i}}$
referring to additional diffusion velocities. The thick arrows (red
online) indicate how the mixing of velocity fields $\overline{v_{i}}$
and $\overline{u_{j}}$, with $i\neq j$, produces the terms proportional
to $\sin\varphi$ in Eq.~\eqref{eq:3.12}. Note that angles such
as $\varphi$ in the schematic symbolize the actual phase angles between
any two vectors, and must therefore be understood as emerging out
of the interaction of all the velocity/momentum vectors involved in
a particular point in space. So, the schematic depicts the totality
of all the ``excited'' velocity/momentum fields involved, labeled
separately for each channel. Upon actual superposition of the two
channels in $x-\textrm{space}$, i.e., in the course of the interactions
among all these field excitations, one then obtains the final intensity
distributions and, according to Eq.~\eqref{eq:3.13}, the particle
trajectories whose emergent behaviour turns out to be characterized
by a ``no crossing'' rule (see Fig.~\ref{fig:2}). In other words,
there is no simple linear superposition between the corresponding
processes for each separate channel 1 and 2, respectively, but a complex
evolution of all the involved sub-quantum processes that leads to
the emergent interference pattern and the associated trajectories.\label{fig:1}}
\end{figure}

We now focus on the specific role of the velocity fields $\VEC u$,
which were present in Section~\ref{sec:path} only implicitly in
the expressions \eqref{eq:3.0a} and \eqref{eq:3.2}. To describe
the required details, each path $i$ be occupied by a Gaussian wave
packet with a ``forward'' momentum $\VEC p_{i}=\hbar\VEC k_{i}=m\VEC v_{i}$.
Moreover, due to the stochastic process of path excitation, the latter
has to be represented also by a large number $N$ of consecutive Brownian
shifts, $\VEC p_{u,\alpha}=m\VEC u_{\alpha}$. Recalling \eqref{eq:3.6},
one obtains for the case of interference at a double slit the total
averaged velocity field (with indices $i=1$ or $2$ referring to
the two slits) 
\begin{equation}
\meanx{\VEC v}_{{\rm tot}}=\meanx{\VEC v}_{{\rm tot},1}+\meanx{\VEC v}_{{\rm tot},2}:=\meanx{\VEC v}_{1}+\frac{\meanx{\VEC u}_{1+}}{2}+\frac{\meanx{\VEC u}_{1-}}{2}+\meanx{\VEC v}_{2}+\frac{\meanx{\VEC u}_{2+}}{2}+\frac{\meanx{\VEC u}_{2-}}{2}.\label{eq:3.8}
\end{equation}
 With two Gaussian distributions $P_{1}=R_{1}^{2}$ and $P_{2}=R_{2}^{2}$
as given in the previous Chapter, one has the corollary of \eqref{eq:3.0-1},
i.e., 
\begin{equation}
R_{{\rm tot}}\meanx{\VEC v}_{{\rm tot}}=R_{1}\meanx{\VEC v}_{{\rm tot},1}+R_{2}\meanx{\VEC v}_{{\rm tot},2}\;,\label{eq:3.9}
\end{equation}
and thus 
\begin{equation}
R_{{\rm tot}}=\left[R_{1}\left(\meanx{\VEC v}_{1}+\frac{\meanx{\VEC u}_{1+}}{2}+\frac{\meanx{\VEC u}_{1-}}{2}\right)+R_{2}\left(\meanx{\VEC v}_{2}+\frac{\meanx{\VEC u}_{2+}}{2}+\frac{\meanx{\VEC u}_{2-}}{2}\right)\right]\frac{\hat{\VEC v}_{{\rm tot}}}{|\meanx{\VEC v}_{{\rm tot}}|}.\label{eq:3.10}
\end{equation}
 To help with the bookkeeping, the schematic of Fig.~\ref{fig:1}
displays all the relevant vectors and some of the corresponding phase
angles. 
Taking into account the conservation of the ``particle momentum''
in both channels, we have $|\meanx{\VEC v}_{{\rm tot}}|=|\meanx{\VEC v}_{{\rm tot},1}|=|\meanx{\VEC v}_{{\rm tot},2}|$,
leading to 
\begin{equation}
R_{{\rm tot}}=\left[R_{1}\left(\hat{\VEC v}_{1}+\frac{\hat{\VEC u}_{1+}}{2}+\frac{\hat{\VEC u}_{1-}}{2}\right)+R_{2}\left(\hat{\VEC v}_{2}+\frac{\hat{\VEC u}_{2+}}{2}+\frac{\hat{\VEC u}_{2-}}{2}\right)\right]\hat{\VEC v}_{{\rm tot}}.\label{eq:3.10a}
\end{equation}
 With Eq.~\eqref{eq:2.7a}, and setting $\VEC u_{1+}\to\VEC u_{1}$
and $\VEC u_{1-}\to-\VEC u_{1}$, one obtains the total average current
\begin{equation}
\small\begin{array}{r@{}r@{\;}c@{\;}r@{\;}c@{\;}r@{\;}l@{}}
\multicolumn{7}{l}{\meanx{\VEC J}_{{\rm tot}}=R_{{\rm tot}}^{2}\meanx{\VEC v}_{{\rm tot}}=R_{1}^{2}\meanx{\VEC v}_{1}+R_{2}^{2}\meanx{\VEC v}_{2}}\\[1.5ex]
+R_{1}R_{2}\biggl\{ & {\displaystyle \left(\meanx{\VEC v}_{1}+\meanx{\VEC v}_{2}\right)\cos(\hat{\VEC v}_{1},\hat{\VEC v}_{2})} & + & {\displaystyle \left(\meanx{\VEC v}_{1}+\frac{\meanx{\VEC u}_{2}}{2}\right)\cos(\hat{\VEC v}_{1},\hat{\VEC u}_{2})} & - & {\displaystyle \left(\meanx{\VEC v}_{1}-\frac{\meanx{\VEC u}_{2}}{2}\right)\cos(\hat{\VEC v}_{1},\hat{\VEC u}_{2})} & \biggr.\\[2ex]
+ & {\displaystyle \left(\frac{\meanx{\VEC u}_{1}}{2}+\meanx{\VEC v}_{2}\right)\cos(\hat{\VEC u}_{1},\hat{\VEC v}_{2})} & - & {\displaystyle \left(-\frac{\meanx{\VEC u}_{1}}{2}+\meanx{\VEC v}_{2}\right)\cos(\hat{\VEC u}_{1},\hat{\VEC v}_{2})} & + & {\displaystyle \left(\frac{\meanx{\VEC u}_{1}}{2}+\frac{\meanx{\VEC u}_{2}}{2}\right)\cos(\hat{\VEC u}_{1},\hat{\VEC u}_{2})} & \biggl.\biggr.\\[2ex]
- & {\displaystyle \left(\frac{\meanx{\VEC u}_{1}}{2}-\frac{\meanx{\VEC u}_{2}}{2}\right)\cos(\hat{\VEC u}_{1},\hat{\VEC u}_{2})} & - & {\displaystyle \left(-\frac{\meanx{\VEC u}_{1}}{2}+\frac{\meanx{\VEC u}_{2}}{2}\right)\cos(\hat{\VEC u}_{1},\hat{\VEC u}_{2})} & + & {\displaystyle \left(-\frac{\meanx{\VEC u}_{1}}{2}-\frac{\meanx{\VEC u}_{2}}{2}\right)\cos(\hat{\VEC u}_{1},\hat{\VEC u}_{2})} & \biggl.\biggr\},
\end{array}\label{eq:3.11}
\end{equation}
 and thus finally 
\begin{equation}
\meanx{\VEC J}_{{\rm tot}}=P_{1}\meanx{\VEC v}_{1}+P_{2}\meanx{\VEC v}_{2}+\sqrt{P_{1}P_{2}}\left(\meanx{\VEC v}_{1}+\meanx{\VEC v}_{2}\right)\cos\varphi+\sqrt{P_{1}P_{2}}\left(\meanx{\VEC u}_{1}-\meanx{\VEC u}_{2}\right)\sin\varphi.\label{eq:3.12}
\end{equation}
Note that Eq.~\eqref{eq:3.12}, upon the identification of $\meanx{\VEC u}_{i}=-\frac{\hbar}{m}\frac{\nabla R_{i}}{R_{i}}$
from Eq.~\eqref{eq:2.7} and with $P_{i}=R_{i}^{2}$, turns out to
be in perfect agreement with a comparable ``Bohmian'' derivation~\cite{Holland.1993,Sanz.2008trajectory}.
The formula for the averaged particle trajectories, then, simply results
from Eq.~\eqref{eq:3.11}, i.e., 
\begin{equation}
\meanx{\VEC v}_{{\rm tot}}=\frac{\meanx{\VEC J}_{{\rm tot}}}{P_{\textrm{tot}}}=\frac{P_{1}\meanx{\VEC v}_{1}+P_{2}\meanx{\VEC v}_{2}+\sqrt{P_{1}P_{2}}\left(\meanx{\VEC v}_{1}+\meanx{\VEC v}_{2}\right)\cos\varphi+\sqrt{P_{1}P_{2}}\left(\meanx{\VEC u}_{1}-\meanx{\VEC u}_{2}\right)\sin\varphi}{P_{1}+P_{2}+2\sqrt{P_{1}P_{2}}\cos\varphi}\;.\label{eq:3.13}
\end{equation}

In Fig.~\ref{fig:2} one can observe a basic characteristic of the
(averaged) particle trajectories, which, just because of the averaging,
are identical with the Bohmian trajectories. In particular, due to
the ``no crossing'' rule for Bohmian trajectories, the particles
coming from, say, the right slit (and expected at the left part of
the screen if a presumed ``classical'' momentum conservation should
hold) actually arrive at the right part of the screen (and \textit{vice
versa} for the other slit). In our sub-quantum approach an explanation
of the ``no crossing'' rule is actually a consequence of a detailed
microscopic momentum conservation. As can be seen in Fig.~\ref{fig:2},
the (Bohmian) trajectories are repelled from the central symmetry
line. However, in our case this is only implicitly due to a ``quantum
potential'', but actually due to the identification of the latter
with a kinetic (rather than a potential) energy: As has already been
stressed in \cite{Groessing.2009origin}, it is the ``heat of the
compressed vacuum'' that accumulates along said symmetry line (i.e.,
as reservoir of ``outward'' oriented kinetic energy) and therefore
repels the trajectories. 

The trajectories in Fig.~\ref{fig:2} exactly obey Eq.~\eqref{eq:3.13}
for the two Gaussian slits shown. The interference hyperbolas for
the maxima characterize the regions where the phase difference $\varphi=2n\pi$,
and those with the minima lie at $\varphi=(2n+1)\pi$, $n=0,1,2,\ldots$
Note in particular the ``kinks'' of trajectories moving from the
center-oriented side of one relative maximum to cross over to join
more central (relative) maxima. In addition to the full accordance
with the trajectories obtained from the Bohmian approach (see~\cite{Bohm.1993undivided},
\cite{Holland.1993}, \cite{Sanz.2008trajectory} and \cite{Sanz.2009context},
for example), in our classical explanation of interference a detailed
``micro-causal'' account of the corresponding kinematics can be
given: Firstly, we note that the last term in Eq.~\eqref{eq:3.12},
which is responsible for the genuinely ``quantum'' behaviour, determines
the movement towards the symmetry line. This term is characterized
by the product of the vector $\meanx{\VEC u}_{1}-\meanx{\VEC u}_{2}$
and $\sin\varphi$, the combined effect of which results in the kinks
typical for Bohmian trajectories (Fig.~\ref{fig:2}). 

Thus, in the cases where the trajectories come from a relative maximum
(bright fringe), the particles loose velocity/momentum in the direction
towards the symmetry line and cross over into the area of the adjacent
relative minimum (dark fringe). From there, they gain velocity/momentum
in the direction towards the symmetry line and thus align with the
other trajectories of the next bright fringe. In other words, one
obtains areas where part of the current (along a relative maximum)
is being removed (``depletion''), or where parts of currents flow
together to produce a newly formed bright fringe (``accumulation''),
respectively. This is in accordance with our earlier description of
quantum interference, where the effects of diffusion wave fields were
explicitly described by alternating zones of heat accumulation or
depletion, respectively \cite{Groessing.2009origin}. Towards the
central symmetry line, then, one observes heat accumulation from both
sides, and due to big momentum kicks from the central accumulation
of heat energy, the forward particle velocities' directions align
parallel to the symmetry axis. With the crossing-over of particle
trajectories being governed by the last, diffusion-related, term on
the right hand side of Eq.~\eqref{eq:3.12}, one finds that for $\varphi=0$
the resulting diffusive current is zero and thus, as total result
of the overall kinematics, no crossing is possible. Further, we note
that our results are also in agreement with the recently published
experimental results by Kocsis~et~al.~\cite{Kocsis.2011observing}.
Here we just comment that, as opposed to the Bohmian interpretation,
we give a micro-causal explanation of these results solely on classical
grounds.

\section{The meaning of the relative phase: Quantum superposition, modular
momentum and the nonlocality of the path excitation field\label{sec:schrodinger}}

Although we have obtained the usual quantum mechanical results, we
have so far not used the quantum mechanical formalism in any way.
However, upon employment of the Madelung transformation for each path
$j$ ($j=1$ or $2$), 
\begin{equation}
\Psi_{j}=R\e^{iS_{j}/\hbar},\label{eq:3.14}
\end{equation}
 and thus $P_{j}=R_{j}^{2}=|\Psi_{j}|^{2}=\Psi_{j}^{*}\Psi_{j}$,
with the definitions \eqref{eq:2.7} and $v_{j}:=\nabla S_{j}/m$,
$\varphi=(S_{1}-S_{2})/\hbar$, and recalling the usual trigonometric
identities such as $\cos\varphi=\frac{1}{2}\left(\e^{i\varphi}+\e^{-i\varphi}\right)$,
etc., one can rewrite the total average current \eqref{eq:3.12} immediately
as 
\begin{equation}
\begin{array}{rl}
{\displaystyle \meanx{\VEC J}_{{\rm tot}}} & =P_{{\rm tot}}\meanx{\VEC v}_{{\rm tot}}\\[3ex]
 & ={\displaystyle (\Psi_{1}+\Psi_{2})^{*}(\Psi_{1}+\Psi_{2})\frac{1}{2}\left[\frac{1}{m}\left(-i\hbar\frac{\nabla(\Psi_{1}+\Psi_{2})}{(\Psi_{1}+\Psi_{2})}\right)+\frac{1}{m}\left(i\hbar\frac{\nabla(\Psi_{1}+\Psi_{2})^{*}}{(\Psi_{1}+\Psi_{2})^{*}}\right)\right]}\\[3ex]
 & ={\displaystyle -\frac{i\hbar}{2m}\left[\Psi^{*}\nabla\Psi-\Psi\nabla\Psi^{*}\right]={\displaystyle \frac{1}{m}{\rm Re}\left\{ \Psi^{*}(-i\hbar\nabla)\Psi\right\} ,}}
\end{array}\label{eq:3.18}
\end{equation}
 where $P_{{\rm tot}}=|\Psi_{1}+\Psi_{2}|^{2}=:|\Psi|^{2}$. The last
two expressions of \eqref{eq:3.18} are the exact well-known formulations
of the quantum mechanical probability current, here obtained without
any quantum mechanics, but just by a re-formulation of \eqref{eq:3.12}.
In fact, it is a simple exercise to insert the wave functions \eqref{eq:3.14}
into \eqref{eq:3.18} to re-obtain \eqref{eq:3.12}.

It is important to note that while the total wave-function $\Psi=\Psi_{1}+\Psi_{2}$
obviously obeys the quantum mechanical superposition principle, no
linear superposition principle holds for our total current ${\displaystyle \meanx{\VEC J}_{{\rm tot}}}$,
as can easily be seen from Eq.~\eqref{eq:3.12}. In accordance with
't Hooft's arguments \cite{t_Hooft.2012quantum}, we have thus demonstrated
with an explicit model how the quantum mechanical superposition principle
is only a calculatory means to describe the effects of sub-quantum
processes, which are actually to be understood as complex behaviours
of ontological microscopic states. This scenario is reassuring, as
the constructed linearity of quantum mechanics appears to be ``sandwiched''
between the sub-quantum and the classical macro-levels, respectively,
i.e., levels where the superposition principle does not hold. So,
applying linear superposition to states other than those constructed
for practical purposes only would lead to wrong predictions about
the behaviours of ontological states.%
\footnote{We thus completely agree with 't Hooft's very basic statement: ``Quantum
wave functions were introduced for the convenience of the computations;
linearity came as a handy tool for making calculations, but it so
happens that quantum superpositions of ontological states themselves
do not describe any real world, and this, as it turns out now, explains
why we do not see quantum mechanical superpositions occuring in the
macro world.'' \cite{t_Hooft.2012quantum}%
} 

Now that the identity between the versions of our classically derived
expression for the average current and the quantum one is established,
we have to confront the claim that it was impossible to reproduce
quantum results with a classical wave theory, because the meaning
of a quantum phase apparently was very different from the meaning
of a classical phase. This claim is detailed particularly clearly
in \cite{Tollaksen.2009quantum}, where the authors look for a mechanism
to explain how the particle, say, at the right, ``knows'' what is
happening at the left slit (i.e., whether the latter is closed or
open, for example). Their explanation of interference from the single-particle
perspective is based on non-local Heisenberg equations of motion for
``modular variables'' like the modular momentum, for example. With
$p$ denoting the usual momentum, the modular momentum is defined
as $p_{mod}:=p\:\textrm{mod}\:\frac{h}{d}=p-n\frac{h}{d}$, where
$d$ is the distance between the slits. As $p\,\mathrm{mod\,\frac{h}{d}}$
has the topology of a circle, nothing changes if in the equations
one replaces $p$ with $p-n\frac{h}{d}$. The main argument now is
that whereas the ordinary momentum (as well as all its higher moments)
is independent of the relative phase between $\Psi_{1}$ and $\Psi_{2}$,
the modular momentum is not. In other words, the authors claim that
the sensitivity of the modular momentum towards changes in the relative
phase makes it impossible to apply classical intuitions to double
slit interference: As the relative phase is a truly non-local feature
of quantum mechanics, they claim, classical notions would have to
fail to describe the essence of interference. Physically, therefore,
it is a non-local effect of having an open or closed slit to produce
a shift in the particle's modular momentum while the expectation values
of its ordinary momentum remain unaffected. 

However, as we have shown, we have an identity in the outcomes of
our classical calculations with the quantum ones. How can this come
about? The answer is clearly given by the fact that the path excitation
field is, by its very definition, a non-local field. Now, one may
argue that non-local fields should have nothing to do with ``classical''
physics, but as we have repeatedly stressed, nothing speaks\emph{
a priori} against the assumption of some non-locally distributed zero-point
oscillations in a purely classical context. Once this is accepted,
the rest follows straightforwardly: instead of using the language
of a ``rotation in the space of a modular variable'', like the ``non-local
exchange of modular momentum'', for example, we have discussed and
described interference with the non-local path excitation field, where
the angle between two unit vectors representing the respective velocity
fields originating from the two slits is given by the relative phase.
As in \cite{Tollaksen.2009quantum} the parameter relevant to describe
the effects of an open or a closed slit, respectively, is given by
the distance between the slits appearing in the modular momentum approach,
so does this same distance also appear in our description, i.e., it
is explicitly given in the formula for the relative phase $\varphi$
in Eq.~\eqref{eq:3.2}. Thus, the appearance of the distances $X$
in our expression for $\varphi$ essentially demonstrates the latter
to be a non-local one. For example, the closing of one slit at either
$-X$ or $+X$ has an immediate effect in that for $v_{x}=0$ the
last term of the relative phase~\eqref{eq:3.2} is changed by a factor
of $\frac{1}{2}$. 

Moreover, if in Eq.~\eqref{eq:3.2} one discards the diffusion-related
term, then $\varphi$ becomes ``classical'' in the usual sense of
the word, and it is only then that the quantum results could not be
reproduced any more. Similarly, in Eq.~\eqref{eq:3.12} it is the
last term proportional to $\sin\varphi$ which determines the genuinely
quantum nature of the whole expression, and it is there where via
the nonlocality of the ``diffusive'' velocities $\mathbf{u_{\textrm{i}}}$
the nonlocality of quantum mechanics becomes manifest. We have thus
shown why our classical approach can produce the results in full accordance
with quantum mechanics. A more detailed discussion of nonlocality
and entanglement within our scheme will be the subject of a paper
in preparation.

\section{Conclusions\label{sec:conclusion}}

We have introduced a quantum as an emergent system by considering
``particles'' as oscillators (``bouncers'') coupling to regular
oscillations of the ``vacuum's'' zero-point field, which they also
generate. Among other features, the dynamics between the oscillator
and the ``bath'' of its thermal environment can be made responsible
not only for Gaussian diffraction at a single slit \cite{Groessing.2010emergence},
but also for the well-known interference effects at a double slit
\cite{Groessing.2012doubleslit,Schwabl.2012quantum}. We have also
shown how the model entails the existence of a path excitation field,
i.e., a field spanned by the average velocity fields $\meanx{\VEC v}(\VEC x,t)$
and $\meanx{\VEC u}(\VEC x,t)$, respectively, where the latter refer
to diffusion processes reflecting also the stochastic parts of the
zero-point field. We have derived, on the basis of classical physics
only, the exact intensity distribution at a screen behind a double
slit, as well as the details of the more complicated particle current,
or of the Bohmian particle trajectories, respectively. In a simple
computer simulation, we have modeled quantum interference with simple
classical rules employing well-known techniques to model diffusion
processes \cite{Mesa_Pascasio.2012classical}.

Moreover, we have refuted claims about the impossibility to model
quantum interference with any classical (and thus also our) model.
The decisive feature of said claims, apparently without a classical,
or ``natural'', explanation, is the non-local effect of opening
or closing one slit on a particle going through the other slit, an
effect which manifests itself in a changed expression for the relative
phase. However, we were able to show explicitly within our classical
approach that the path excitation field in this case must change to
produce exactly the same effect on the relative phase. This both qualifies
as a truly non-local effect within our approach and provides the identity
with the usual quantum mechanical predictions.

Finally, upon comparison with the usual quantum mechanical formalism,
we have demonstrated with our explicit model how the quantum mechanical
superposition principle is only a calculatory means using non-ontological
wave-functions to describe the effects of sub-quantum processes, which
are actually to be understood as complex, nonlinear behaviours of
ontological microscopic states.


\providecommand{\href}[2]{#2}\begingroup\raggedright\endgroup

\end{document}